\documentclass[10pt,conference]{IEEEtran}
\IEEEoverridecommandlockouts

\usepackage{cite}
\usepackage{algorithmic}
\usepackage{graphicx}
\usepackage{textcomp}
\usepackage{xcolor}
\usepackage{graphicx}
\usepackage{subfiles}
\usepackage{enumitem}
\usepackage{adjustbox}
\usepackage{float}
\usepackage{tabularx}
\usepackage{multirow}
\usepackage{caption}
\usepackage{amsmath,amsfonts}
\usepackage{url}
\usepackage{booktabs}
\usepackage{siunitx}
\usepackage{fancyvrb}
\usepackage{rotating}
\usepackage{array}
\usepackage{collcell}
\usepackage[many]{tcolorbox}
\usepackage{todonotes}
\usepackage{makecell}
\usepackage{soul}
\usepackage{listings}
\usepackage{balance}

\newtcolorbox{boxK} {
    sharpish corners,
    boxrule = 0pt,
    toprule = 3pt,
    enhanced,
    fuzzy shadow = {0pt}{-2pt}{-0.5pt}{0.5pt}{black!35},
    boxsep=0mm,
    left=2mm,
    right=2mm,
    top=1.5mm,
    bottom=1.5mm
}

\definecolor{diffgreen}{rgb}{0.1,0.7,0.2}
\lstdefinestyle{javastyle}{
    language=java,
    basicstyle=\ttfamily\footnotesize,
    breaklines=true,
    commentstyle=\color{gray},
    keywordstyle=\color{blue},
    stringstyle=\color{orange},
    showspaces=false,
    showstringspaces=false,
    showtabs=false,
    tabsize=2,
    frame=none,
    xleftmargin=1mm,
    moredelim=[is][\color{diffgreen}]{@g@}{@g@}
}

\definecolor{commandblue}{rgb}{0.2,0.4,0.8}
\definecolor{paramgreen}{rgb}{0.2,0.6,0.2}
\lstdefinestyle{bashstyle}{
    language=bash,
    basicstyle=\ttfamily\footnotesize,
    breaklines=true,
    commentstyle=\color{gray},
    keywordstyle=\color{commandblue},
    stringstyle=\color{paramgreen},
    showspaces=false,
    showstringspaces=false,
    showtabs=false,
    tabsize=2,
    frame=none,
    xleftmargin=1mm,
    moredelim=[is][\color{commandblue}]{@c@}{@c@},
    moredelim=[is][\color{paramgreen}]{@p@}{@p@}
}

\newcommand{\customcell}[1]{%
  \begin{tabular}[t]{@{}l@{}r@{}}
    #1
  \end{tabular}%
}

\newcolumntype{C}{>{\collectcell\customcell}c<{\endcollectcell}}

\def\BibTeX{{\rm B\kern-.05em{\sc i\kern-.025em b}\kern-.08em
    T\kern-.1667em\lower.7ex\hbox{E}\kern-.125emX}}

\newcommand\numSurvey{40}

\title{
\textbf{\huge{From Technical Excellence to Practical Adoption: Lessons Learned Building an ML-Enhanced Trace Analysis Tool}}
}

\author{
\IEEEauthorblockN{Kaveh Shahedi\IEEEauthorrefmark{1}, 
Matthew Khouzam\IEEEauthorrefmark{2}, 
Heng Li\IEEEauthorrefmark{1}, 
Maxime Lamothe\IEEEauthorrefmark{1}, 
Foutse Khomh\IEEEauthorrefmark{1}}
\IEEEauthorblockA{\IEEEauthorrefmark{1}Polytechnique Montréal, Montréal, Canada}
\IEEEauthorblockA{\IEEEauthorrefmark{2}Ericsson AB, Montréal, Canada}
}

\begin{document}

\maketitle

\begin{abstract}
System tracing has become essential for understanding complex software behavior in modern systems, yet sophisticated trace analysis tools face significant adoption gaps in industrial settings. Through a year-long collaboration with Ericsson Montréal, developing TMLL (Trace-Server Machine Learning Library, now in the Eclipse Foundation), we investigated barriers to trace analysis adoption. Contrary to assumptions about complexity or automation needs, practitioners struggled with translating expert knowledge into actionable insights, integrating analysis into their workflows, and trusting automated results they could not validate. We identified what we called the \emph{Excellence Paradox}: technical excellence can actively impede adoption when conflicting with usability, transparency, and practitioner trust. TMLL addresses this through adoption-focused design that embeds expert knowledge in interfaces, provides transparent explanations, and enables incremental adoption. Validation through Ericsson's experts' feedback, Eclipse Foundation's integration, and a survey of 40 industry and academic professionals revealed consistent patterns: survey results showed that 77.5\% prioritize quality and trust in results over technical sophistication, while 67.5\% prefer semi-automated analysis with user control, findings supported by qualitative feedback from industrial collaboration and external peer review. Results validate three core principles: cognitive compatibility, embedded expertise, and transparency-based trust. This challenges conventional capability-focused tool development, demonstrating that sustainable adoption requires reorientation toward adoption-focused design with actionable implications for automated software engineering tools.
\end{abstract}

\section{Introduction}
\label{introduction}
Software systems operating across distributed architectures and multi-core processors demand sophisticated performance analysis capabilities. System tracing has emerged as a powerful technique for capturing detailed runtime behavior, providing detailed visibility into system execution~\cite{sigelman2010dapper,zhou2022industrial, gebai2018survey}. While automated software engineering has revolutionized many aspects of software development, from code generation to testing, the automation of dynamic performance analysis remains a critical gap in industrial practice~\cite{zhao2023large,zhang2022machine,fraser2014large}. Despite sophisticated tracing tools~\cite{sigelman2010dapper,fonseca2007xtrace, desnoyers2006lttng,kimuftrace, perf, gperftools, kaldor2017canopy} and established analytical techniques~\cite{cornelissen2009systematic, ardelean2018performance, huang2021tprof}, practitioners struggle to leverage trace data effectively for performance analysis in industrial settings.

This paper presents our systematic investigation of trace analysis adoption barriers through a year-long industrial collaboration with Ericsson Montréal and the development of TMLL (Trace Server Machine Learning Library)\footnote{https://github.com/eclipse-tracecompass/tmll}. Rather than a purely technical contribution, we focus on empirical insights about the misalignment between academic assumptions regarding automated tool adoption and the reality faced by practitioners in industrial environments. TMLL's integration into the Eclipse Foundation as an official project provides external validation of our industry-focused approach.

\begin{figure}[]
\centering
\includegraphics[width=0.95\columnwidth]{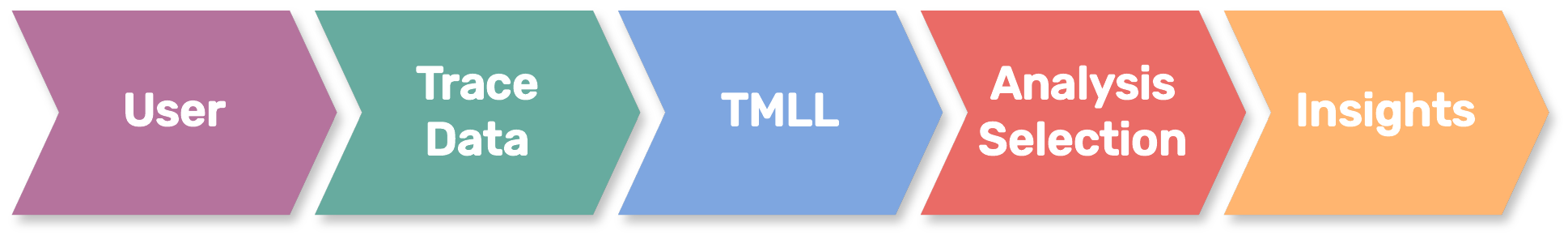}
\captionsetup{justification=centering}
\caption{TMLL workflow: from trace data import to actionable insights through automated analysis selection}
\vspace{-5mm}
\label{fig:tmll-quick-overview}
\end{figure}

Our research journey began with seemingly reasonable hypotheses grounded in conventional automated software engineering wisdom: practitioners were not adopting trace analysis tools because existing tools were too complex, required excessive manual effort, and lacked sufficient automation. We hypothesized that democratizing sophisticated analytical techniques through machine learning would naturally lead to widespread adoption, following established patterns in other automated software engineering domains~\cite{li2024ai,davis1989perceived,memon2017google,hilton2016ci}. Through sustained industrial collaboration and systematic practitioner engagement, these assumptions proved fundamentally flawed.

Our empirical investigation revealed that real adoption barriers were categorically different from our initial assumptions. Practitioners struggled not with tool complexity per se, but with three distinct automation challenges: (1) translating expert analytical knowledge into actionable automated insights, (2) integrating automated analysis capabilities into existing development workflows without disrupting established practices, and (3) building confidence in automated results that they could otherwise not independently validate. Most significantly, we discovered what we term the \emph{Excellence Paradox:} technical excellence can actively hinder adoption when it comes at the cost of usability, transparency, and practitioner trust.

Through iterative development, continuous stakeholder engagement, and systematic analysis of both successful and failed approaches, our research addresses three key questions:

\textit{\textbf{RQ1}: What specific adoption barriers do practitioners face when using sophisticated trace analysis tools in industrial settings, and how do these barriers manifest in practice?}

\textit{\textbf{RQ2}: Which design principles enable TMLL to achieve sustainable adoption in real-world development workflows within our study context?}

\textit{\textbf{RQ3}: What gaps exist between academic tool development assumptions and industrial adoption requirements, as revealed through sustained practitioner feedback?}

We validated our understanding through multiple convergent methods: a comprehensive survey of {\numSurvey} industry practitioners and academic researchers, TMLL's integration into the Eclipse Foundation as an official project, early adoption experiences across multiple users within Ericsson and external organizations, and sustained engagement from practitioners throughout the development process.

The primary contributions of this work are empirical insights from our industrial collaboration that suggest four directions for automated trace analytical tool development: (1) an empirical characterization of adoption barriers, with documenting gaps between theoretical automated trace analysis capabilities and practical adoption barriers, (2) a design methodology for adoption-focused automation tools derived from our iterative approach, (3) a validation framework showing how sustained practitioner feedback reveals misalignments between academic assumptions and industrial needs, and (4) documented guidance for researchers building automated trace analysis tools for complex technical domains, emphasizing cognitive efficiency alongside computational sophistication.

\section{Background and Related Work}
\label{related-works}
Our work builds upon research spanning two key domains: trace analysis tools and systems, and machine learning-based trace and log analysis. We structure this section to provide the necessary background while positioning our contributions within the broader research landscape.

\subsection{Trace Analysis Tools and Systems}

System tracing is a fundamental technique for understanding complex software behavior in modern distributed systems. Distributed tracing systems capture the flow of requests across multiple services, enabling performance analysis and debugging of microservice architectures~\cite{gebai2018survey,desnoyers2006lttng,gperftools,fenlason1988gnu,kimuftrace,perf, huang2021tprof, shahedi2023tracing, ardelean2018performance,ezzati2021debugging}. Pioneered by foundational work, including Google's Dapper~\cite{sigelman2010dapper}, which established the architectural principles used by virtually all contemporary tracing platforms, and X-Trace~\cite{fonseca2007xtrace}, which introduced end-to-end tracing across administrative domains with metadata propagation.

Systems like Magpie~\cite{barham2004magpie} and Canopy~\cite{kaldor2017canopy} demonstrated the feasibility of automatic request flow extraction and large-scale trace processing, with Canopy processing over 1 billion traces daily in Facebook's production environment. Modern distributed tracing has evolved significantly with systems like DeepFlow~\cite{shen2023deepflow} and TraceWeaver~\cite{ashok2024traceweaver}, which eliminate the need for application instrumentation through eBPF-based network monitoring and statistical timing analysis, respectively.

Advanced sampling techniques are critical for managing trace data volume at scale. Traditional uniform random sampling proves inadequate for capturing rare but important events, leading to sophisticated approaches like Sifter~\cite{lascasas2019sifter}, which uses machine learning to bias sampling toward edge-case code paths, and retroactive sampling systems~\cite{zhang2023benefit} that capture detailed traces only when problems are detected. Performance analysis frameworks such as The Mystery Machine~\cite{chow2014mystery} have demonstrated automatic causal model construction from component logs, enabling critical path analysis and bottleneck identification in distributed systems.

Compared to these existing systems, TMLL addresses a fundamentally different challenge. While tools like Dapper and Jaeger focus on trace collection and visualization, and systems like DeepFlow automate trace generation, TMLL specifically targets the interpretation gap between raw trace data and actionable performance insights. Unlike Canopy's focus on large-scale trace processing infrastructure, TMLL emphasizes cognitive compatibility and workflow integration to achieve practitioner adoption.

\subsection{Machine Learning-based Trace and Log Analysis}

Machine learning applications in trace and log analysis have evolved from simple statistical anomaly detection to sophisticated deep learning approaches. The field encompasses log parsing and preprocessing~\cite{zhu2019tools, he2018identifying}, where automated methods extract structured information from unstructured log data, and anomaly detection techniques that identify unusual system behavior patterns.

Deep learning approaches have shown particular promise for trace analysis. DeepLog~\cite{du2017deeplog} pioneered LSTM-based sequence modeling for log anomaly detection with workflow construction for root cause analysis. LogAnomaly~\cite{meng2019loganomaly} extended this using template2vec semantic embeddings combined with LSTM networks to detect both behavioral and quantitative anomalies in unstructured logs. Recent work addresses practical deployment challenges, with LogRobust~\cite{zhang2019robust} using ensemble methods and transfer learning for evolving log formats, and comprehensive evaluations~\cite{le2022log} revealing significant gaps between academic results and practical deployment scenarios.

Microservice-specific approaches have emerged to handle the unique challenges of distributed trace analysis. ServiceAnomaly~\cite{panahandeh2023serviceanomaly} combines distributed traces with profiling metrics using Context Propagation Graphs for anomaly detection in microservice systems. For high-performance computing environments, tools like STAT~\cite{arnold2007stat} provide scalable debugging through tree-based analysis and process equivalence classes. Root cause analysis has advanced through ensemble ML methods~\cite{lin2018predicting} and automated performance debugging techniques~\cite{liu2013autoanalyzer} that combine multiple algorithmic approaches for comprehensive system analysis.

However, despite technical advances in ML-based analysis, adoption remains limited in industrial settings. Most existing work focuses on algorithmic improvements rather than addressing practical barriers preventing practitioners from effectively using these sophisticated analytical capabilities. Empirical studies~\cite{le2022log} confirm significant challenges in real-world deployment scenarios, including concept drift, training data requirements, and interpretability concerns.

TMLL differs from existing ML-based trace analysis tools in its explicit focus on adoption barriers rather than algorithmic sophistication. While systems like DeepLog and LogAnomaly optimize for detection accuracy using complex deep learning models, TMLL prioritizes explainable results and embedded expertise with cognitive compatibility for existing practitioner workflows. TMLL addresses the gap identified by Le and Zhang~\cite{le2022log} between academic ML approaches and practical industrial deployment through its adoption-focused design methodology.

\section{Methodology}
\label{sec:methodology}
\subsection{Research Design and Approach}
This study employs a participatory action research methodology~\cite{davison2004action, Siew2013ParticipatoryAR} within an industrial setting to systematically investigate trace analysis adoption barriers and develop evidence-based solutions through iterative tool development. Our approach combines sustained practitioner engagement with empirical development, enabling simultaneous study of adoption challenges and validation of proposed solutions in real industrial contexts. The research follows a mixed-methods design integrating qualitative stakeholder feedback with quantitative validation data, specifically targeting the deployment of intelligent trace analysis tools in production environments.

\subsection{Research Context and Setting}
Our research was conducted through collaboration with Ericsson Montréal, a major telecommunications company operating complex distributed systems requiring sophisticated performance analysis. The organization's multi-service architectures with intricate subsystem interactions provided an ideal setting for investigating trace analysis adoption barriers, as performance degradation could have a significant operational and business impact. This industrial context featured characteristics essential for our study: technically sophisticated users, clear performance requirements, substantial trace datasets, and strong organizational incentives for performance optimization.

The core collaboration centered on a 5-member trace analysis team with over 15 years of experience working on tracing for high-performance systems, serving as the primary technical partner. Additionally, we engaged with several product development teams (5-15 members each) responsible for telecommunications infrastructure components requiring sophisticated performance analysis. This industrial partnership provided authentic validation of ML-enhanced trace analysis approaches in production environments.

\begin{figure*}[htbp]
\centering
\includegraphics[width=0.85\textwidth]{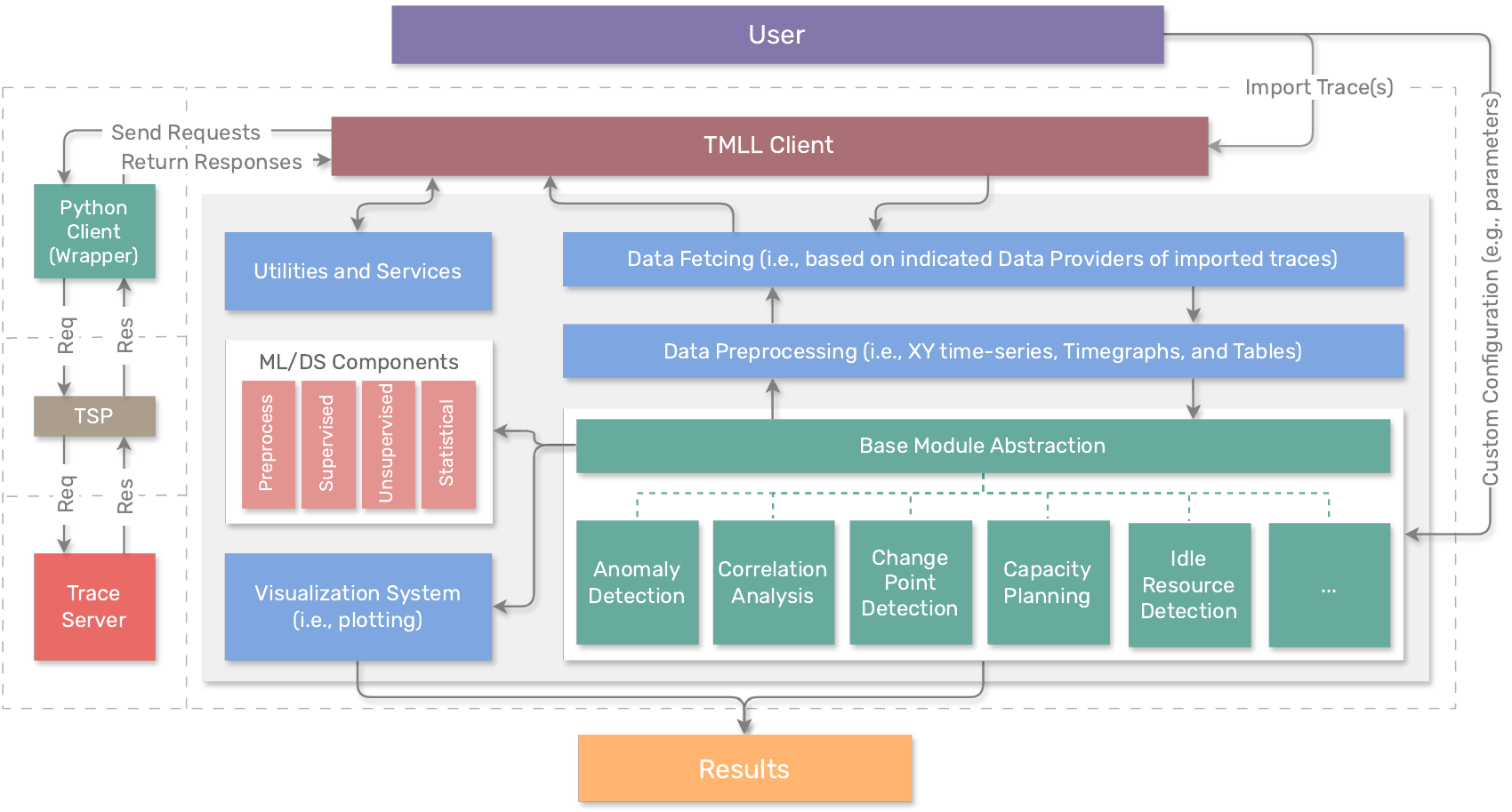}
\captionsetup{justification=centering}
\caption{TMLL's architecture: data processing, analysis modules, and visualization components with client-server integration}
\label{fig:overview}
\end{figure*}

TMLL was developed as an extension to Trace Compass~\cite{tracecompass}, an established trace-analysis tool trusted by thousands of users, leveraging established protocols for optimized handling of large-scale trace data. As shown in Figure~\ref{fig:overview}, The system architecture consists of five primary layers: (1) client interface layer providing Python-based access, (2) data processing layer implementing intelligent data fetching and preprocessing, (3) base module abstraction layer enabling consistent interfaces, (4) specialized analysis modules implementing machine learning techniques, and (5) visualization and results layer. The tool enables practitioners to import trace files into TMLL's client, create experiments from these traces, and then apply various ML-enhanced analysis modules (anomaly detection, correlation analysis, change point detection, etc.) to gain insights from the trace data. The adoption of TMLL by the Eclipse Foundation as an official project validates the industrial relevance and technical soundness of our approach.

\subsection{Research Timeline and Phases}
The research was conducted over approximately 12 months (May 2024 - April 2025), organized into four phases aligned with industrial development cycles: \textbf{Phase 1} (Months 1-3): Problem investigation and infrastructure design through stakeholder interviews and proof-of-concept development. \textbf{Phase 2} (Months 4-6): Core module development, including trace-based anomaly detection, correlation analysis, and memory leak analysis, with continuous feedback integration. \textbf{Phase 3} (Months 7-9): Infrastructure improvements and reliability enhancements based on early adoption feedback, with early adoption beginning in Month 7. \textbf{Phase 4} (Months 10-12): Integration capabilities development (including workflow and deployment capabilities), formal evaluation, and successful Eclipse Foundation submission. 

\subsection{Participant Recruitment and Data Collection}
Our data collection strategy employed two complementary approaches: sustained practitioner engagement throughout development and systematic validation using a structured survey, designed to capture both depth and breadth of industrial adoption patterns.

\textbf{Development Phase Engagement:} Throughout TMLL development, we maintained continuous engagement with approximately 25-30 practitioners using purposive sampling~\cite{palinkas2015purposive} within Ericsson and extended professional networks. Participants represented diverse industrial roles, including software developers, system performance specialists, DevOps engineers, and academic researchers. This sustained engagement provided iterative feedback on design decisions, adoption barriers, and usage patterns through biweekly progress presentations, monthly formal meetings, and semi-structured informal discussions focused on practical deployment challenges.

\textbf{Validation Survey Design and Execution:} To validate insights beyond our development context, we designed a mixed-methods survey with 15 Likert-scale questions measuring adoption preferences and barriers, and 8 open-ended questions exploring practitioner experiences and priorities. We distributed the survey through organizational networks, professional contacts, and academic communities, using snowball sampling~\cite{biernacki1981snowball} to reach practitioners across organizations. The survey recruited 40 practitioners from multiple companies and academic researchers, providing validation across diverse industrial contexts.

\begin{figure*}[htbp]
\centering
\includegraphics[width=\textwidth]{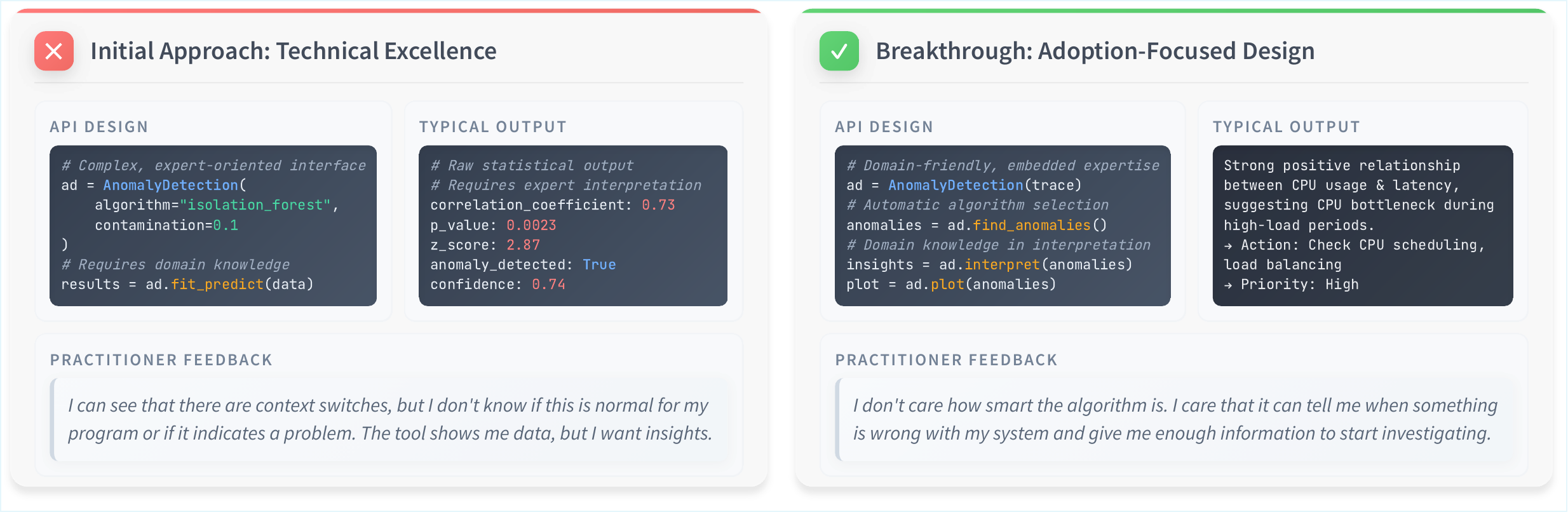}
\captionsetup{justification=centering}
\caption{The Excellence Paradox: When Technical Sophistication Hinders Adoption. TMLL's design philosophy prioritizes practitioner cognitive compatibility over algorithmic sophistication, demonstrated through our development at Ericsson.}
\label{fig:excellence_paradox}
\end{figure*}

\subsection{Tool Development and Analysis}
TMLL implements a modular architecture designed to support incremental adoption in industrial environments. The base module abstraction provides consistent interfaces for data processing, analysis execution, and result visualization, enabling practitioners to integrate capabilities progressively without disrupting existing workflows. Our development methodology emphasized rapid prototyping with continuous validation through feature-driven development and comprehensive documentation suitable for enterprise deployment.

The iterative development approach enabled continuous refinement based on practitioner feedback, with each development cycle validated through direct usage in Ericsson's production environment. This industrial validation approach ensures that design decisions reflect real-world deployment constraints rather than theoretical considerations.

\subsection{Research Hypotheses}
Based on our preliminary investigation and existing literature on developer tool adoption~\cite{davis1989perceived,li2024ai}, we formulated three core hypotheses about adoption-focused tool design for intelligent trace analysis systems:

\textit{$\mathcal{H}1$ (Cognitive Compatibility):} Tools that augment existing practitioner mental models achieve higher adoption than those requiring new analytical paradigms.

\textit{$\mathcal{H}2$} Practitioners prefer tools that embed expert knowledge in interfaces rather than expose analytical capabilities for manual configuration.

\textit{$\mathcal{H}3$ (Transparency-Trust Relationship):} Explainable automated ML-enhanced analysis achieves higher practitioner trust and adoption than black-box sophisticated algorithms.

These hypotheses guide our design decisions and evaluation methodology, providing testable predictions about the relationship between interface design choices and practitioner adoption patterns in industrial trace analysis environments.

\section{Adoption Challenges in Trace Analysis}
\label{sec:challenges}
Our empirical investigation through sustained practitioner engagement across 5 teams at Ericsson, with 12 observing and 12 months of development, identified three fundamental barriers preventing trace analysis tool adoption in industrial practice. These challenges represent systematic failures in how traditional tools approach the gap between analytical capabilities and practitioner needs in production environments, ultimately impacting the productivity and effectiveness of trace analysis in telecommunications and other enterprise domains.

\subsection{Expertise Distribution and Knowledge Transfer Barriers}

Effective trace analysis requires specialized knowledge to interpret patterns, correlate indicators, and distinguish normal system behavior from problematic conditions~\cite{he2018identifying,cornelissen2009systematic,Pennington1987}. In industrial environments, this expertise is concentrated among a few practitioners (Nearly none of the teams have dedicated performance specialists), creating significant bottlenecks when teams need trace analysis capabilities for production incident response and performance optimization.

Most practitioners can collect traces using standard tools (FTrace, DTrace, LTTng, perf) but struggle to extract actionable insights from raw data. Our observational studies at Ericsson revealed significant interpretation gaps: practitioners could identify obvious anomalies (CPU spikes $>$ 90\%) but missed subtle patterns indicating developing problems that could lead to production failures. Expert analysts consistently recognized meaningful correlations that non-experts failed to detect, even when provided with identical visualization tools~\cite{Pennington1987,Goncales2021}. For example, experts identified 85 \% of performance regression cases that non-experts missed entirely during controlled evaluation sessions.

The business impact of these expertise bottlenecks extends beyond immediate technical challenges. Teams without dedicated performance specialists face increased mean time to resolution for performance issues, higher operational costs, and reduced system reliability. In telecommunications environments like Ericsson's, where system performance directly affects service quality and customer satisfaction, these expertise gaps represent significant organizational risk.

\subsection{Scale, Complexity, and Workflow Integration Challenges}

Modern distributed systems generate massive trace volumes with hundreds of simultaneous metrics across multiple subsystems~\cite{zhou2021fault,gan2019deathstarbench, ardelean2018performance, kaldor2017canopy}. Production systems at Ericsson typically generate between 100,000 and 150,000 events per second across thousands of concurrent processes, requiring correlation of temporal patterns across CPU usage, memory allocation, I/O operations, network traffic, and application-level events. Manual correlation analysis becomes intractable beyond simple scenarios involving around 50 metrics over 15-minute time windows.

Practitioners reported spending significant time analyzing curated datasets with numerous metrics, only to conclude they couldn't determine whether observed patterns represented problems or normal system variation. The combinatorial explosion of potential correlations ($N\times(N-1)/2$ combinations for N metrics) overwhelms human analytical capacity, particularly when temporal offsets and non-linear relationships must be considered~\cite{Sweller1988,Goncales2021}. For instance, real-world traces with 50 simultaneous metrics create 1,225 potential pairwise correlations, far exceeding human processing capabilities.

Performance issues often manifest through subtle interactions across multiple trace dimensions rather than obvious single-metric anomalies~\cite{zhou2021fault,du2017deeplog, liao2021locating, chow2014mystery}. Performance regressions typically involve a sensible increase across multiple metrics rather than dramatic spikes in individual measurements. Practitioners struggle to maintain awareness of multi-dimensional patterns while examining individual metric trends.

Existing trace analysis tools exacerbate these challenges by requiring practitioners to learn new analytical paradigms rather than fitting into existing mental models. Practitioners conceptualize performance problems in terms of system-level behaviors (``application becomes slow'', ``memory usage grows over time'') rather than statistical abstractions (correlation coefficients, anomaly scores)~\cite{he2018identifying,Pennington1987,cornelissen2009systematic}, yet traditional tools present results in statistical terms requiring translation into actionable insights.

Interviews with more than 40 practitioners revealed consistent preferences for tools that \emph{``tell me what's wrong and what to do about it''} rather than \emph{``show me the data and let me figure it out''}. This represents a fundamental mismatch between tool capabilities and practitioner cognitive models. Integration failures occur not due to technical incompatibility but because tools don't align with how practitioners conceptualize performance problems~\cite{davis1989perceived,BanoZowghi2015, sadowski2018cacm}. Successful integration requires workflow compatibility rather than just technical interoperability~\cite{li2024ai,davis1989perceived}.

Traditional trace analysis requires practitioners to switch between familiar debugging contexts (application logs, system monitors) and unfamiliar analytical contexts (statistical software, specialized visualization tools). This context switching introduces considerable time overhead per analysis session and reduces practitioner willingness to engage with trace analysis regularly, ultimately limiting the effectiveness of performance monitoring and optimization efforts.

\subsection{Empirical Evidence from Industrial Deployment}

Our field studies at Ericsson provided quantitative evidence for these challenges in production. Teams with dedicated performance specialists resolved most of the performance issues within hours, while teams without specialists required at least 2-3x of time on average for similar issues. This performance gap translates directly to operational costs and service availability impact in telecommunications environments.

Tool adoption patterns suggest that sophisticated analytical platforms tend to have a lower adoption rate compared to simpler visualization tools, which remain in use despite their limited analytical capabilities.
This data confirms that adoption barriers extend beyond technical sophistication to fundamental usability and workflow integration issues that affect long-term tool viability in industrial settings.

Our survey of {\numSurvey} practitioners confirmed these challenges across diverse organizational contexts beyond Ericsson. Most significantly, 77.5\% identified ``quality and trust in results'' as a primary adoption barrier, followed closely by 75.0\% citing ``learning curve and required expertise,'' while ``integration with existing tools'' ranked third at 55.0\%. This ranking validates that adoption barriers extend beyond technical capabilities to fundamental transparency and trust issues, with practitioners prioritizing result reliability over workflow compatibility across different industrial domains. The elevation of trust concerns above integration requirements reinforces our core finding that technical excellence without explainability actively hinders adoption.

\section{TMLL: Adoption-Focused Architecture for Industrial Deployment}
\label{sec:architecture}

This section presents TMLL's architecture as a systematic response to the adoption barriers identified in Section~\ref{sec:challenges}. Rather than optimizing for algorithmic sophistication, TMLL prioritizes cognitive compatibility, workflow integration, and embedded expertise to achieve sustained adoption, with each architectural decision addressing specific challenges from our collaboration with Ericsson.

\subsection{Adoption-Driven Architectural Philosophy}

TMLL's architecture embodies what we term \textbf{adoption-first design}, where each layer and component is designed to minimize adoption barriers rather than maximize technical capabilities. Figure~\ref{fig:excellence_paradox} illustrates the fundamental shift from technical excellence to adoption-focused design that enabled TMLL's successful engagements. This philosophy emerged from observing that practitioners consistently abandoned technically sophisticated tools in favor of simple, familiar approaches that fit their existing mental models and workflows.

The architecture addresses three adoption requirements: \textbf{expertise accessibility} (i.e., minimizing the need for specialized knowledge), \textbf{cognitive compatibility} (i.e., fitting existing practitioner mental models), and \textbf{workflow integration} (i.e., enhancing rather than replacing established practices). These requirements drove fundamental architectural decisions, including embedded expertise in interfaces, transparent explanation generation, and modular capability introduction that enables incremental adoption without workflow disruption.

\subsection{Three-Layer Architecture for Adoption Success}

TMLL implements a three-layer architecture where each layer specifically addresses adoption barriers identified in our industrial collaboration. The architecture maintains technical rigor necessary for production trace analysis while prioritizing practitioner accessibility and workflow compatibility over algorithmic sophistication.

\subsubsection{Standardized Data Processing Layer: Eliminating Infrastructure Barriers}
TMLL's foundational layer (i.e., Trace Server) automatically transforms heterogeneous trace formats (e.g., FTrace, DTrace, LTTng, custom formats) into a standardized representation (i.e., Pandas data frames), removing the need for practitioner data cleaning expertise that has traditionally limited trace analysis to specialists.

The ``Data Preprocessing'' class implements comprehensive error recovery mechanisms for real-world trace data quality issues, automatically handling missing values (i.e., over than 60 \% of 
"unacalibrated" traces contain gaps such as lost events, after tuning, tracers will not have temporal gaps, but may have informational gaps ), format inconsistencies (i.e., fields assigned with a:=b, a=b, a:b, a=[b,c] and timestamps from multiple timezones stored without geolocation), and temporal alignment problems across concurrent data streams.

This layer eliminates a major adoption barrier commonly observed across many organizations and teams, including Ericsson: practitioners often abandon trace analysis when faced with complex data preprocessing requirements. Performance optimizations include streaming processing for traces up to terrabytes of data and parallelization~\cite{tracecompass}, ensuring that the abstraction layer doesn't compromise the performance requirements of production environments. This design principle (i.e., hiding complexity without sacrificing capability) proved essential for adoption in technical environments where both ease of use and production-scale performance are mandatory.

\subsubsection{Analysis Engine Layer: Embedded Expertise for Democratized Analysis}
TMLL hides complex statistics behind one uniform API, letting practitioners ``click and learn'' instead of ``tune and pray.''  
Its six built-in modules cover the trace questions teams hit most often:
\begin{itemize}[leftmargin=*]
  \item \textbf{Anomaly Detection}: scans any desired metric, picks an appropriate detector on the fly (e.g., z-score, IQR, Isolation Forest), and finds suspicious timestamps worth a closer look.
  \item \textbf{Memory Leak Detection}: follows pointer lifetimes (e.g., \texttt{malloc}/\texttt{free}) and memory growth trends to flag leaks (e.g., \texttt{malloc} without \texttt{free}) and finds original call sites.
  \item \textbf{Correlation Analysis}: maps which metrics rise or fall together (and in what order) to speed causal reasoning.
  \item \textbf{Change Point Detection}: spots hidden shifts in system behavior by voting across multiple metrics aggregately.
  \item \textbf{Capacity Planning}: projects metrics' upcoming behavior (e.g., CPU, memory, disk, network), warns when defined thresholds will be crossed, and suggests scaling actions.
  \item \textbf{Idle Resource Identification}: detects long idle stretches of metrics (e.g., underutilized memory, imbalance CPU core workloads) and recommends consolidation or re-balancing.
\end{itemize}

\begin{figure}[]
\centering
\includegraphics[width=\columnwidth]{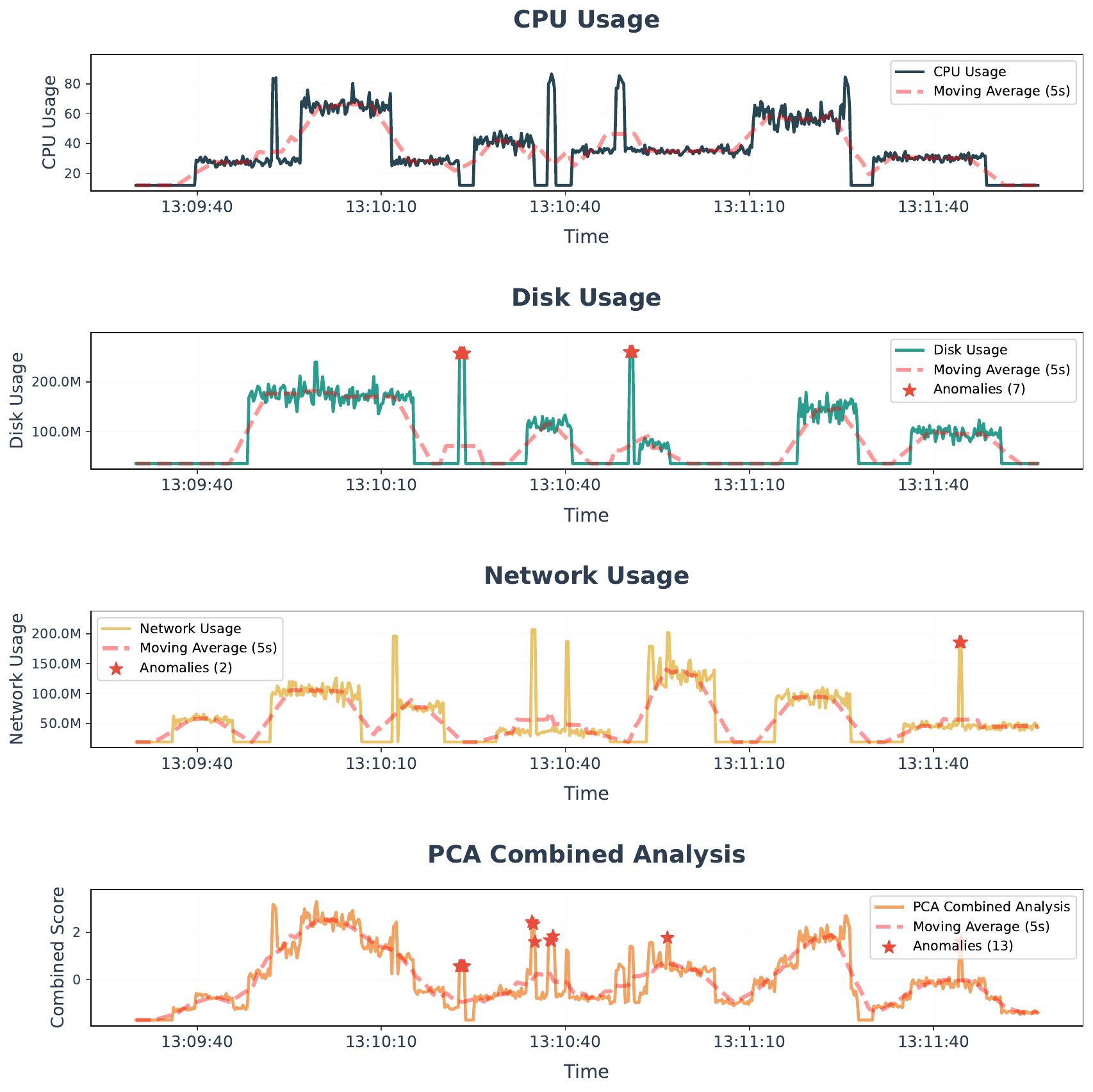}
\captionsetup{justification=centering}
\caption{Multi-metric anomaly detection output showing CPU, disk, and network usage analysis with identified anomalies and combined PCA scoring}
\label{fig:example-anomaly}
\end{figure}

\begin{figure}[]
\centering
\includegraphics[width=0.9\columnwidth]{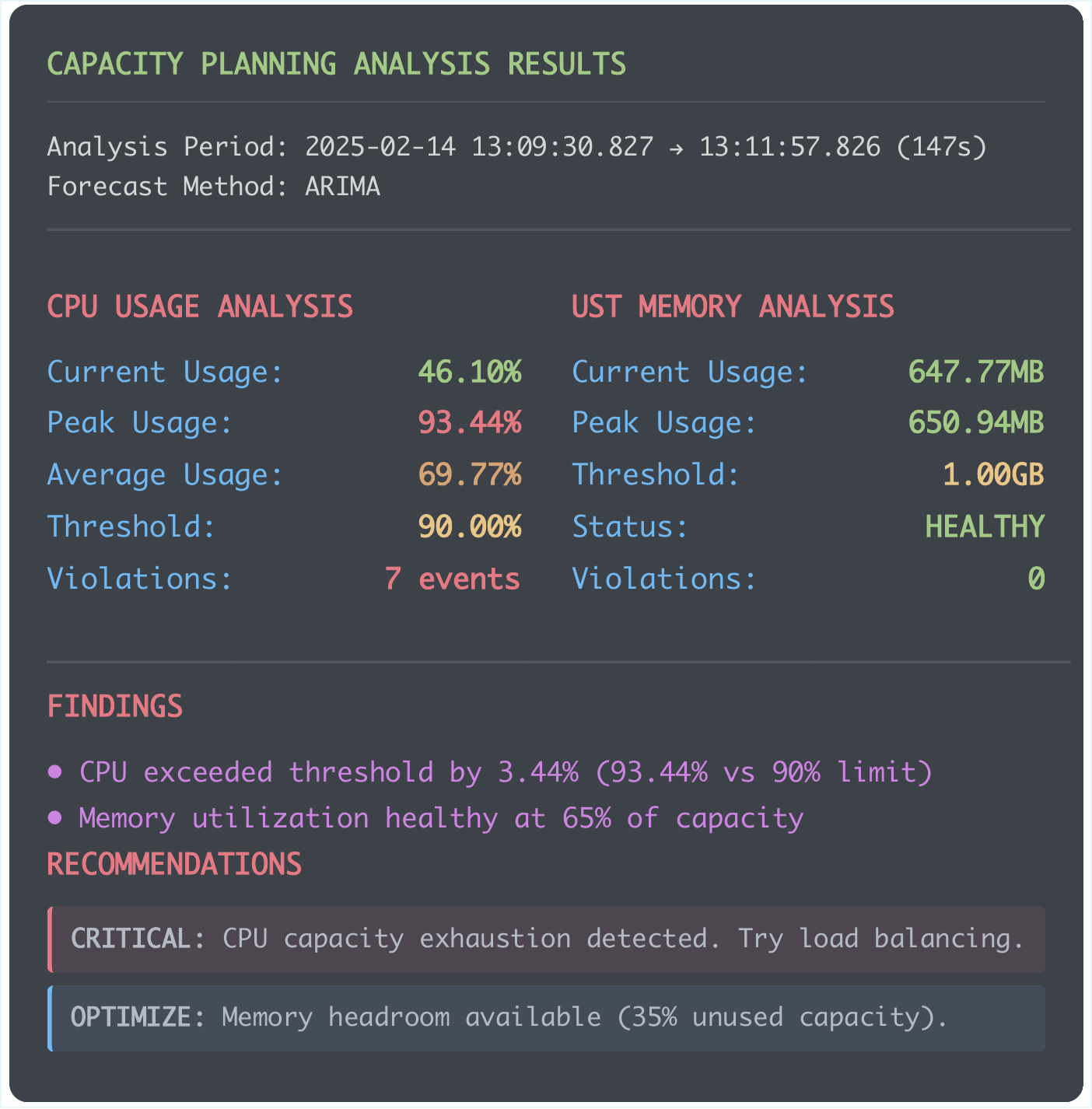}
\captionsetup{justification=centering}
\caption{TMLL capacity planning analysis results with threshold violations, resource utilization assessment, and actionable recommendations}
\label{fig:example-capacity}
\end{figure}

Figures~\ref{fig:example-anomaly} and~\ref{fig:example-capacity} illustrate representative examples of TMLL's analysis output, demonstrating how the tool translates complex analytical results into actionable insights with clear visualizations and domain-relevant recommendations.

TMLL's standardized, extensible architecture proved essential during deployment, where all modules extend a shared abstract base class to ensure consistent data access, processing, analysis, and output. This design allows teams to create tailored analyses, ranging from simple calculations to sophisticated analyses, without compromising usability or workflow integration. By presenting results in domain-relevant, actionable terms through a unified interface, TMLL eliminates the need for practitioners to learn different configurations, supporting scalable adoption across diverse teams and validating the value of a consistent, flexible framework.

\subsubsection{Interface and Interpretation Layer: Cognitive Compatibility and Workflow Integration}

The presentation layer addresses the workflow integration challenge by maintaining cognitive compatibility with existing performance engineering practices~\cite{Pennington1987,cornelissen2009systematic}. The ``Visualization'' and ``Document Generation'' classes translate analytical results into actionable insights that align with practitioner mental models rather than requiring adoption of new visualization paradigms.

This layer emerged from the critical insight that adoption failures often result from interface mismatches rather than algorithmic limitations. Practitioners at Ericsson consistently preferred tools that augmented their existing debugging approaches over comprehensive analytical platforms requiring new paradigms. Results include domain-relevant explanations, confidence assessments, and recommended actions rather than statistical outputs, bridging the gap between sophisticated analysis and practical decision-making.

The interface supports various visualization formats (e.g., XY line graphs, heatmaps, box plots) and exports to standard formats for integration with existing toolchains, ensuring that TMLL enhances rather than replaces established workflows. This integration capability proved essential for sustained adoption, as practitioners could gradually incorporate TMLL's insights into their familiar debugging processes.

\subsection{Production-Scale Implementation and Deployment}

TMLL's implementation emphasizes production readiness and enterprise adoption through a carefully designed technology stack and deployment architecture. The system is built in Python using established scientific computing libraries including NumPy~\cite{harris2020array}, SciPy~\cite{jones2001scipy}, and pandas~\cite{mckinney2010data} for numerical analysis, with Matplotlib~\cite{Hunter2007matplotlib} and Seaborn~\cite{Waskom2021} providing visualization capabilities. Machine learning functionality is supported through scikit-learn~\cite{Pedregosa2011scikit-learn}, TensorFlow~\cite{AbaBar16Tensorflow}, and Statsmodels~\cite{seabold2010statsmodels}, while the framework accommodates custom library integration to meet specialized analytical requirements.

The deployment architecture scales from individual developer workstations to high-performance computing clusters, enabling flexible usage patterns that range from personal debugging sessions to enterprise-wide performance monitoring. This architectural flexibility proved critical during Ericsson's adoption, where TMLL supported both individual practitioner workflows and large-scale organizational analytics initiatives.

Central to TMLL's extensibility is the Base Module abstraction, which provides a standardized interface for custom analytical development~\cite{RunesonHost2009}. This design enables practitioners to implement domain-specific analyses within TMLL's infrastructure while maintaining consistency across data access patterns, result formatting, and visualization generation. The module system supports multiple analysis types through unified interfaces, allowing users to extend TMLL's capabilities without requiring them to learn new architectural paradigms.

The technology stack deliberately avoids exotic dependencies that could complicate enterprise deployment, instead prioritizing proven libraries that balance analytical power with operational simplicity~\cite{li2024ai,davis1989perceived}. This approach ensures that TMLL can be deployed across diverse enterprise environments while maintaining the analytical sophistication required for complex performance analysis tasks.

\section{Validation and Results}
\label{sec:validation}
\begin{table*}[htbp]
\centering
\captionsetup{justification=centering}
\caption{Practitioner Priorities vs. Initial Research Assumptions Survey Analysis (n=40)}
\label{tab:survey_insights}
\footnotesize
\setlength{\tabcolsep}{4pt}
\renewcommand{\arraystretch}{1}
\color{black}
\begin{tabular}{
    @{}
    >{\raggedright\arraybackslash}p{0.30\textwidth}
    >{\centering\arraybackslash}p{0.12\textwidth}
    >{\raggedright\arraybackslash}p{0.48\textwidth}
    @{}
}
\toprule
\textbf{Capability/Feature} & 
\textbf{\% Valued} & 
\textbf{Key Insight} \\
\midrule
\multicolumn{3}{c}{\textbf{Most Valued Capabilities}} \\
\cmidrule{1-3}
Anomaly Detection & 77.5\% & Strong preference for actionable insights over algorithmic sophistication \\
\cmidrule{2-3}
Root Cause Analysis & 75.0\% & Contextual interpretation valued more than correlation detection \\
\cmidrule{2-3}
Data in Python-ready Formats & 42.5\% & Infrastructure important but secondary to analytical insights \\
\midrule
\multicolumn{3}{c}{\textbf{Automation vs. Control Preferences}} \\
\cmidrule{1-3}
Fully Automated Analysis & 20.0\% & \multirow{2}{0.48\textwidth}{Overwhelming preference for guided analysis with transparency over black-box automation} \\
\cmidrule{2-2}
Semi-automated with User Control & 67.5\% & \\
\midrule
\multicolumn{3}{c}{\textbf{Primary Adoption Factors (Ranked by Priority)}} \\
\cmidrule{1-3}
Quality and Trust in Results & 77.5\% & Explainability and reliability now top priority for adoption \\
\cmidrule{2-3}
Learning Curve and Expertise Required & 75.0\% & Critical need for embedded expertise, not simplified tools \\
\cmidrule{2-3}
Integration with Existing Tools & 55.0\% & Workflow compatibility important but secondary to trust \\
\cmidrule{2-3}
Documentation and Support & 42.5\% & Essential infrastructure for enterprise adoption \\
\midrule
\multicolumn{3}{c}{\textbf{Feature Importance (Rated 4-5/5)}} \\
\cmidrule{1-3}
Automated Data Preprocessing & 100.0\% & Universal agreement: infrastructure work as high-value innovation \\
\cmidrule{2-3}
Interactive Visualizations & 100.0\% & Exploration and discovery more valued than automated conclusions \\
\cmidrule{2-3}
Anomaly Detection Capabilities & 100.0\% & Automated pattern recognition essential for practical adoption \\
\cmidrule{2-3}
Root Cause Analysis Features & 100.0\% & Diagnostic capabilities critical for production environments \\
\cmidrule{2-3}
Extensibility for Custom Modules & 97.5\% & Near-universal need for adaptation to local requirements \\
\midrule
\multicolumn{3}{c}{\textbf{Adoption Likelihood}} \\
\cmidrule{1-3}
Likely/Very Likely to Try TMLL & 85.0\% & Strong validation of adoption-focused design approach \\
\bottomrule
\end{tabular}
\end{table*}

We validated TMLL's adoption-focused design through systematic evaluation combining iterative development feedback from industrial collaboration at Ericsson and multiple external organizations, quantitative survey data from diverse organizations, and external peer review through Eclipse Foundation acceptance. This section presents validation methodology, results from development collaboration, and formal hypothesis testing that demonstrates the broader applicability of our approach.

\subsection{Multi-Method Validation Strategy}
Our validation approach assessed TMLL's design principles across diverse contexts through three complementary methods: sustained development collaboration with Ericsson Montréal practitioners, cross-organizational survey validation, and external peer review.

\subsubsection{Development Collaboration Validation at Ericsson}
The primary validation occurred through sustained development collaboration at Ericsson Montréal over 12 months, involving iterative feedback from cross-functional development teams spanning diverse engineering domains, including radio baseband, DevOps, AI, and infrastructure teams. This industrial collaboration provided authentic validation of ML-enhanced trace analysis design approaches through continuous practitioner engagement across multiple technical disciplines during tool development.

Rather than post-deployment metrics, validation came through iterative design cycles where Ericsson practitioners provided feedback on each development phase. This approach enabled us to validate design decisions in real-time, adjusting TMLL's architecture and capabilities based on practitioner responses to prototypes, feature demonstrations, and hands-on evaluation sessions.

The iterative feedback process validated our core design principles through practitioner reactions to different interface approaches, algorithm presentations, and workflow integration strategies. Comments like \textit{``I want the insights that an expert would have, but I don't want to become an expert myself''} from practitioners confirmed that our embedded expertise approach resonated with the actual needs of industrial users.

\subsubsection{Cross-Organizational Survey Design and Methodology}
To validate insights beyond Ericsson's organizational context, we developed a mixed-methods instrument with 15 quantitative ratings (5-point Likert scales) and 8 qualitative feedback questions structured around three research questions derived from our development collaboration experience. The survey instrument included attention checks and consistency verification across related questions to ensure response validity~\cite{RunesonHost2009}.

We recruited {\numSurvey} practitioners through organizational networks, professional contacts, and academic communities using snowball sampling~\cite{biernacki1981snowball}. Participants included software developers (55.0\%, n=22), researchers (12.5\%, n=5), team managers/leads (7.5\%, n=3), DevOps/SRE engineers (5.0\%, n=2), students (12.5\%, n=5), performance engineers (2.5\%, n=1), data scientists (2.5\%, n=1), and trace product managers (2.5\%, n=1) across different organizations. This diverse sample provided validation across different organizational contexts and technical domains beyond telecommunications.

\subsection{Development Collaboration Insights and Design Validation}
\subsubsection{Iterative Feedback Patterns and Design Evolution}
The development collaboration at Ericsson revealed consistent patterns in practitioner responses that validated our adoption-focused design approach. Early prototypes focusing on algorithmic sophistication received lukewarm responses, while interfaces emphasizing embedded expertise and domain-relevant interpretation generated strong positive feedback.

Practitioners consistently preferred demonstration scenarios where TMLL provided immediate actionable insights over those requiring interpretation of statistical outputs. This pattern emerged across multiple feedback sessions and validated our shift from capability-focused to adoption-focused design. The feedback process confirmed that technical excellence alone was insufficient for practitioner acceptance.

Incremental development proved essential for validation, as practitioners could evaluate each design decision in context rather than responding to abstract concepts. This iterative validation process enabled us to identify and address adoption barriers before they became embedded in the tool architecture.

\subsubsection{Workflow Integration Validation Through Development Feedback}
Ericsson practitioners' feedback validated our workflow integration approach through their responses to different interface designs and integration strategies. Rather than requiring comprehensive workflow changes, TMLL's design to augment familiar approaches received consistently positive feedback during development sessions. Notably, as TMLL is implemented in Python, many participants, particularly those from AI teams with limited experience in traditional trace-based performance engineering tools, can understand and engage with the tool much more effectively.
Practitioners emphasized that tools should ``fit well into existing debugging practices'' without disrupting established responsibilities. This feedback pattern validated our cognitive compatibility design principle and influenced architecture decisions that prioritize integration over replacement of existing practices.

\subsection{Survey Validation of Design Principles}
\subsubsection{Practitioner Preferences and Adoption Factor Validation}
Survey results, shown in Table~\ref {tab:survey_insights}, strongly validated our adoption-focused design approach across diverse organizational contexts. Most significantly, quality and trust in results emerged as the top adoption priority (77.5\%, n=31), followed closely by manageable learning curves (75.0\%, n=30), while integration with existing tools ranked third (55.0\%, n=22). This prioritization shift from our initial assumptions reinforces that practitioners value transparency and reliability above workflow compatibility, strongly supporting our emphasis on explainable analysis over black-box sophistication.

Critical validation came from automation preferences: 67.5\% (n=27) preferred semi‑automated analysis with user control over fully automated solutions (20.0\%, n=8). This indicates that practitioners value access to expert‑level capabilities without the burden of expert‑level knowledge, provided they can maintain transparency and control. These findings confirm our embedded‑expertise approach and validate our core design philosophy that addresses the Excellence Paradox across diverse organizational contexts.

\subsubsection{Feature Value Assessment and Capability Priorities}
As illustrated in Table~\ref{tab:survey_insights}, universal agreement emerged on core infrastructure capabilities: automated preprocessing (100\%), interactive visualizations (100\%), anomaly detection (100\%), and root cause analysis (100\%) were rated as highly valuable (4-5/5), with extensibility achieving near-universal approval (97.5\%). This validates our infrastructure-focused approach that prioritizes practical utility over algorithmic sophistication.

The shift in capability preferences further validates our design approach: anomaly detection (77.5\%) and root cause analysis (75.0\%) were most valued for providing actionable insights rather than algorithmic sophistication. Comments emphasized wanting ``contextual interpretation, not just correlation detection'' and preferring tools that provide ``actionable insights, not algorithmic sophistication.'' This validates our embedded expertise approach that translates sophisticated analysis into domain-relevant insights, confirming that practitioners prioritize interpretable results over complex algorithms.

\subsection{Formal Hypothesis Validation}
Our systematic validation across development collaboration and cross-organizational survey provides quantitative evidence for testing our three research hypotheses about adoption-focused tool design:

\subsubsection{$\mathcal{H}1$ (Cognitive Compatibility): VALIDATED}
Both Ericsson's development feedback and survey results (55.0\% valuing integration with existing tools) confirmed that practitioners prefer tools augmenting existing mental models over those requiring new analytical paradigms. The elevation of quality and trust in results to the top adoption priority (77.5\%) demonstrates that cognitive compatibility extends beyond workflow integration to encompass result interpretability. Survey comments consistently emphasized preference for tools that ``augment existing practices'' rather than ``require learning new approaches.''

\subsubsection{$\mathcal{H}2$ (Embedded Expertise): VALIDATED}
Development collaboration feedback and survey validation (67.5\% preferring guided over black-box approaches, 100\% valuing automated preprocessing and intelligent analysis capabilities) strongly confirmed our embedded expertise approach. The preference for semi-automated analysis (67.5\% vs. 20.0\% for fully automated) demonstrates practitioners' desire for expert-level capabilities without requiring expert-level knowledge.

\subsubsection{$\mathcal{H}3$ (Transparency-Trust): VALIDATED}
Strong preference for semi-automated approaches (67.5\%) over fully automated solutions (20.0\%) shows the importance of transparency. Most significantly, quality and trust in results emerged as the primary adoption factor (77.5\%), the highest-ranked concern among practitioners. Development collaboration feedback consistently emphasized understanding ``how the tool reached its conclusions'', validating that transparency serves as essential infrastructure for tool adoption in expert domains.

\subsection{External Validation By Eclipse Foundation Integration}
The Eclipse Foundation's adoption of TMLL as an official project\footnote{https://projects.eclipse.org/projects/tools.tracecompass.tmll} provided rigorous external validation through peer review by industry experts and academic researchers working on performance analysis tools. This review evaluated both technical merit and broader industry relevance, providing validation beyond our collaboration and survey results.

Review feedback confirmed broader applicability of our design insights: \textit{``The focus on practitioner adoption rather than algorithmic sophistication addresses a real gap in performance tooling''} and \textit{``The emphasis on interpretability and workflow integration reflects challenges we've seen across multiple organizations.''} This external validation demonstrates that TMLL's adoption-focused design principles extend beyond telecommunications to other enterprise domains requiring sophisticated trace analysis capabilities.

\subsection{Key Validation Findings and Design Implications}
Our systematic validation across development collaboration, cross-organizational survey, and external peer review revealed four critical insights with broader implications for enterprise tool development:

\textbf{Expertise Distribution vs. Knowledge Transfer:} Development collaboration feedback and survey data confirmed that practitioners prefer tools embedding expert knowledge rather than facilitating expert knowledge acquisition. The high priority placed on learning curve concerns (75.0\%) validates the democratization of sophisticated capabilities through embedded expertise rather than training programs or documentation.

\textbf{Trust Through Transparency:} Development feedback and survey results (77.5\% citing result trust as the top adoption factor) demonstrated that interpretability serves as essential trust infrastructure for intelligent analysis tools. This elevation of trust above all other concerns shows that transparency enables practitioners to validate conclusions against domain knowledge. This transparency, in turn, builds the confidence necessary for tool acceptance across diverse contexts.

\textbf{Cognitive Compatibility Over Technical Excellence:} Strong preference for semi-automated analysis with user control (67.5\%) and emphasis on manageable learning curves (75.0\%) confirms our design philosophy of cognitive compatibility. While workflow integration ranked lower (55.0\%), the overwhelming preference for transparent, guided analysis over black-box automation demonstrates that successful enterprise tools should augment existing mental models rather than requiring paradigm shifts, regardless of technical superiority.

\textbf{Iterative Development Enables Validation:} The development collaboration approach proved essential for validating design decisions in context rather than through abstract evaluation. This methodology enabled the identification and resolution of adoption barriers before they became embedded in the tool architecture.

The findings show that successful developer tools in expert domains require fundamental reorientation from capability-focused to adoption-focused design, with implications extending beyond trace analysis to other complex analytical domains requiring expert knowledge for effective industry use.

\section{Threats to Validity}
\label{sec:threats-to-validity}
\textbf{Internal Validity:} Participants were recruited through convenience sampling from existing professional networks, potentially limiting perspective diversity and introducing selection bias toward practitioners interested in analytical tools. Practitioner feedback during structured sessions may have been influenced by demonstration contexts and social desirability bias rather than reflecting natural usage patterns.

\textbf{External Validity:} Primary development occurred within a single telecommunications organization, potentially limiting generalizability to other organizational contexts and technical domains. Findings may be specific to telecommunications and high-performance computing environments, requiring validation in domains like web development or embedded systems where trace analysis challenges differ.

\textbf{Construct Validity:} We lacked comprehensive quantitative adoption metrics due to organizational policies, relying primarily on qualitative indicators (satisfaction, feedback, continued usage) rather than objective usage analytics. Our definition of ``successful adoption'' may be influenced by TMLL's developmental state rather than inherent adoption barriers.

\textbf{Reliability:} Some insights relied on informal feedback and observational notes that may affect reproducibility. Qualitative data analysis involved subjective interpretation of practitioner feedback, potentially introducing researcher bias in pattern identification and insight generation.

\section{Conclusion}
\label{sec:conclusion}
This work reports our systematic investigation of trace analysis adoption barriers through a year-long industrial collaboration with Ericsson Montreal and the development of TMLL, revealing the \emph{Excellence Paradox}: technical excellence can hinder adoption when it comes at the cost of usability and trust. Through development collaboration, cross-organizational survey validation (n={\numSurvey}), and external peer review leading to Eclipse Foundation adoption, we validated three key design principles: (1) practitioners prefer embedded expert knowledge over expertise development requirements, (2) cognitive compatibility outweighs technical sophistication for adoption, and (3) transparency and user control build trust in intelligent analysis tools. Our findings provide actionable guidance: prioritize workflow integration over algorithmic advancement, implement modular architectures supporting incremental adoption, ensure analytical transparency, and incorporate adoption metrics in evaluation frameworks. TMLL's Eclipse Foundation integration and strong practitioner interest (85\% willing to try) demonstrate that adoption-focused design can achieve both technical rigor and practical utility, with implications extending beyond trace analysis to other expert domains requiring sophisticated automated tools.

\bibliographystyle{ieeetr}
\balance
\bibliography{tmll}

\end{document}